\documentstyle[12pt]{article}
\def\bea{\begin{eqnarray}}
\def\eea{\end{eqnarray}}

\def\beq{\begin{equation}}
\def\eeq{\end{equation}}
\def\ba{\beq\new\begin{array}{c}}
\def\ea{\end{array}\eeq}
\def\be{\ba}
\def\ee{\ea}

\makeatletter
\newdimen\normalarrayskip              
\newdimen\minarrayskip                 
\normalarrayskip\baselineskip
\minarrayskip\jot
\newif\ifold             \oldtrue            \def\new{\oldfalse}
\def\arraymode{\ifold\relax\else\displaystyle\fi} 
\def\eqnumphantom{\phantom{(\theequation)}}     
\def\@arrayskip{\ifold\baselineskip\z@\lineskip\z@
     \else
     \baselineskip\minarrayskip\lineskip2\minarrayskip\fi}
\def\@arrayclassz{\ifcase \@lastchclass \@acolampacol \or
\@ampacol \or \or \or \@addamp \or
   \@acolampacol \or \@firstampfalse \@acol \fi
\edef\@preamble{\@preamble
  \ifcase \@chnum
     \hfil$\relax\arraymode\@sharp$\hfil
     \or $\relax\arraymode\@sharp$\hfil
     \or \hfil$\relax\arraymode\@sharp$\fi}}
\def\@array[#1]#2{\setbox\@arstrutbox=\hbox{\vrule
     height\arraystretch \ht\strutbox
     depth\arraystretch \dp\strutbox
     width\z@}\@mkpream{#2}\edef\@preamble{\halign
\noexpand\@halignto
\bgroup \tabskip\z@ \@arstrut \@preamble \tabskip\z@ \cr}%
\let\@startpbox\@@startpbox \let\@endpbox\@@endpbox
  \if #1t\vtop \else \if#1b\vbox \else \vcenter \fi\fi
  \bgroup \let\par\relax
  \let\@sharp##\let\protect\relax
  \@arrayskip\@preamble}
\def\eqnarray{\stepcounter{equation}%
              \let\@currentlabel=\theequation
              \global\@eqnswtrue
              \global\@eqcnt\z@
              \tabskip\@centering
              \let\\=\@eqncr
              $$%
 \halign to \displaywidth\bgroup
    \eqnumphantom\@eqnsel\hskip\@centering
    $\displaystyle \tabskip\z@ {##}$%
    \global\@eqcnt\@ne \hskip 2\arraycolsep
         $\displaystyle\arraymode{##}$\hfil
    \global\@eqcnt\tw@ \hskip 2\arraycolsep
         $\displaystyle\tabskip\z@{##}$\hfil
         \tabskip\@centering
    &{##}\tabskip\z@\cr}

\begin{document}

\begin{flushright}
ITEP/TH-80/98\\
hepth/9812250
\end{flushright}
\vspace{0.5cm}
\begin{center}
{\LARGE \bf Integrability of the RG flows and}
\vspace{0.2cm}
{\LARGE \bf the bulk/boundary correspondence}

\bigskip
{\bf A. Gorsky}

\vspace{0.5cm}
\bigskip
{ ITEP, Moscow, 117259, B.Cheryomushkinskaya 25}

\setcounter{footnote}0

\begin{abstract}
We suggest that RG flows in the N=2 SUSY YM theories are governed by the
pair of the integrable systems.
The main dynamical ingredient amounts
from the  interaction of the small size instantons
with the regulator degrees of freedom. The relation with the
bulk/boundary correspondence is discussed.
\end {abstract}
\end{center}
\bigskip
\section{Introduction}
Recently the correspondence between N=4 finite SYM theory and
IIB string theory in $AdS_5 \times S^5$ has been formulated \cite{mal}
which provides the recipe for the calculation
of the quantum correlators in the SYM theory in terms of the
solution of classical equation of motion in $AdS_5 \times S^5$ IIB supergravity
with the proper boundary conditions \cite{gkp,w1}.
To get the link with the models with less amount of SUSY
it is necessary to include into the game some UV scale
since these theories are not finite and require regularization.
Regulator UV scale penetrates into the IR region via the dimensional transmutation
phenomena and provides  the IR $\Lambda$ type scale.
The attempt to introduce
$\Lambda$  scale was carried out in \cite{w2} where
the M5 brane wrapped around the torus has been considered.
The antiperiodic boundary conditions on the torus are imposed
for fermions so the SUSY is broken because of their induced
tree masses. The corresponding geometry was identified as the
"noncritical black hole in $AdS_{7}\times S^4"$
geometry .

In this letter we discuss a scenario based on the
SUSY breaking by the explicit introduction of the regulator masses.
The simplest example of this type is the N=4 theory softly broken
down to N=2 by the mass of the adjoint  hypermultiplet. The
low energy effective action can be
found for the finite and infinite (N=2 case) adjoint masses \cite{sw}.
The solution is described in terms of the dynamical systems like
periodic Toda chain \cite{gkmmm} or Calogero model \cite{dw} . Moreover
the second dynamical system of the Whitham type was identified with
some RG flows in the model \cite{gkmmm,hokri}. Both dynamical systems can be
considered as the effective motion on some moduli spaces; the Toda-Calogero
one - on the nonperturbative hidden
Higgs branch \cite{g1} while the Whitham system - on the Coulomb branch.
However the SYM interpretation of the "time" evolution in the Calogero-Toda
systems was obscure.

We would like to conjecture  that both dynamical systems can be interpreted
as the RG flows. The key point appears to be related with the
nonperturbative instanton degrees of freedom.
The regulator mass cuts the instantons of the small radii but it turns out
that the back reaction of the small instantons on the regulators
has to be taken into account. The
consistency condition for the instanton interactions with
regulators  yields the first dynamical system. The coordinate scale
factor serves as the time variable in the Toda-Calogero systems
$t=logr$ and the
very role of this integrability is to provide the SUSY invariant renormalization
for the nonperturbative effects. Degrees of freedom come from the
instanton moduli space. The second Whitham type dynamics corresponds to the
motion of the branes  which provide the worldvolume
for the d=4 theory. There are different times in Whitham dynamics; the
simplest one is proportional to the Hitchin time while others  can be
considered as the scale factors in the momentum space.

Let us note that the instanton induced
interactions provide a rather complicated picture
at the regulator scale.
The glue for such viewpoint comes from
the embedding of the integrable many-body systems into the KdV-KP type
hierarchy as a special solutions known for a while \cite{poles}.
The Riemann surface describing the low-energy effective
action appears as a kind of the Fermi surface for the regulator fermions and BPS spectrum
allows the alternative interpretation \cite{gorpei}. The very dynamics sounds
from this point of view as the motion of the  vortices on the Fermi surface
as the function of the coordinate scale factor.

Another issue close to this scenario concerns the relation with the
bulk/boundary picture. In the brane picture we immediately arrive at the
AdS spaces since the Dp brane interpolate between the Minkowskian space at
infinity and $AdS_{p+2}\times S^{d-p-2}$ near the horizon. We shall deal with D0 branes
therefore  $AdS_2$ can be expected. It will be shown how the integrable degrees of freedom
acquire the suggestive interpretation in terms of the AdS spaces.
Finally we shall make some
comments on the calculation of the correlators in the N=2 SYM theory in terms
of the Whitham approach \cite{gmmm} along the line of reasoning in
\cite{gkp,w1} treating the spectral curve of the Toda-Calogero type system as
a  solution of the supergravity equation of motion.

Let us make a few comments on the literature. The remarks on the scale
interpretation of the dynamics in the radial
AdS variable can be found in many papers
(see, for instance, \cite{susw} for general discussion and \cite{dz} for some
explicit examples ). The several points concerning  the possible
integrability of RG flows are discussed in \cite{pol}.  The paper is
organized as follows. In section 2 we formulate the conjecture and compare
the different viewpoints concerning the integrability interpretation.  In
section 3 the relation to the bulk/boundary issue is discussed.  Section 4
contains the concluding remarks and speculations.

\section{RG flows versus integrability}

In this section we consider the picture along the radial AdS direction in the
coordinate space. Let us start with the  picture for  N=4 case. The basic example is
the metric of D3 brane
\be
ds^2=H^{-\frac{1}{2}}dx_{||}^2 + H^{\frac{1}{2}}(dr^2+ r^2d\Omega_{5}^2)\\
H=1+ \frac{4\pi gN\alpha^{'2}}{r^4}
\ee
where $X_{||}$ are four coordinates along the D3 worldvolume and
$d\Omega_{5}^2$ is the five-sphere metric.
It clearly interpolates between the Minkowskian asymptotics
and the product of $AdS_5$
and $S^5$. It is convenient
to discuss the position of the instanton along this direction which has been
identified in this language as D-instanton
\cite{bg}. Using the explicit form of the Green
function in the AdS space it is clear from  the form of the instanton solution
for the  dilaton
\be
\exp (\phi)=g_{s}+\frac{24\pi}{N^2} \frac{r^4 r_{in}^4}{(r_{in}^2+(x-x_0)^2)^4}
\ee
that the instanton radius $r_{in}$ is the position of the instanton in the
radial direction.

Consider the case of the broken conformal symmetry. Adjoint hypermultiplet
plays the role of the regulator field and its mass M serves as a cutoff for
the small size instantons.  In other words we can imagine that the instantons
interact with the "regulator brane" located at some UV scale of order
$M^{-1}$ in the radial direction. We suggest that interaction of the small
size instantons with regulators yields the integrable dynamics of the Hitchin
type and the the radial coordinate in AdS space plays the role of time
variable in the integrable dynamics. The rest of the paper can be considered as
an attempt to check  this suggestion from the different viewpoints.

Let us start with a few general comments
concerning the integrals of motion
in the Hitchin dynamics  treating "time" as the coordinate
RG scale. The most naive argument follows from the remark that
the Hamiltonian of the Hitchin type system can be identified with
the component of the N=2 anomaly multiplet which
due to the conformal anomaly generates the dilation.
Secondly it is known that the action variables in the dynamical
system exactly coincide with the
Seiberg-Witten variables $a_{i}$ providing
the spectrum of the "electric" BPS states in the spectrum. Since
the BPS spectrum appears as the central charges in N=2 SUSY
algebra it is RG invariant indeed. This  suggests that the number of
degrees of freedom of the underlying dynamical system has to
coincide with the number of independent central charges in
the theory. One more comment amounts from the fact that all
integrable systems under consideration admit the Lax representation which
is the first order matrix differential
equation. Certainly it is natural from
the RG point of view.

Let us turn now to the description in IIA picture.
We consider the system of
$N_{c}$ D4 branes localized at points
$v_{i}=<\phi_{i}>$ on the complex $v$ plane with one D0 brane per each.
Regulators have to cut off the large
momenta or in other terms large distance region on the $v$ plane.
"Nonperturbative" D0 branes can be connected by
strings with the regulator branes. Hence in the IIA representation we
have localized D4-D0 system and  strings connecting D0 branes with regulators
at the UV scale.

It is instructive to compare
this picture with the one for N=2 SQCD. In the SQCD case
fundamental matter is represented
by D6 branes localized at points $v_{i}=m_{i}$ where $m_{i}$ is the mass of the
i-th fundamental flavour. D6 and D4 branes are connected by strings which
become infinite if we send the quark mass to infinity.
In this limit fundamentals can be considered as the regulator fields.
Due to the
dimensional transmutation $\Lambda$ scale survives
and we see the  agreement
with the smooth transition from the spin chains to Toda chain which is known
in the integrability language.

The link with the integrable dynamics
looks as follows. As it was shown in \cite{g1} in IIA picture it is
convenient to consider the degrees of freedom as the coordinates of D0 branes
along the different directions. To compare it with discussion above note that
we have a string attached at each D0 brane extended to the regulator scale.
The string has the  $x_{6}$ coordinate (the D4 branes are extended in this
direction while D0 are localized) related to  D0`s one and just this
coordinate  roughly was identified with coordinate of the i-th Calogero-Toda
particle. Therefore once again we have the same representation of the
coordinate variable in the dynamical system as the coordinate
of the end of the string
at the regulator brane - now via the holographic projection of the bulk D0
branes.

This kind of the holography amounts to
the relation with the KdV-KP dynamics. It is  known for a long time \cite{poles}
that the elliptic Calogero system is related with the elliptic solutions of
KdV-KP equations \footnote{Recently this correspondence has been generalized
to the 2d Toda-Ruijsenaars relation \cite{kz} that is to 5d theories in the
SYM theory language. }.  The picture derived provides the attractive
explanation of this relation.  Indeed the connection can be formulated in
terms of the elliptic KdV $\tau$ function
\be
\tau (x,t)=\prod S(x-x_{i}(t))
\ee
where S was identified with the Weiershtrass $\sigma$
function. If the $\tau$ function obeys the Hirota equation for KdV(KP) then
$x_{i}$ have to evolve with respect to the corresponding k-th time according
to the Hamiltonian of the Calogero system $H=TrL^k$ where L is the Lax
operator for the Calogero system with the fixed coupling constant.

Consider the simplest  equation for the eigenfunction of the
rational KdV Lax
operator  which has the following form
\be
(\partial_x^2+\sum_{i}\frac{2}{(x-x_{i}(t))^2})\Psi(\lambda,x,t)=
\lambda \Psi(\lambda,x,t).
\ee
It is natural to identify poles on the $x$ plane as the positions of the
vortices created by the ends of the string coming from D0 brane on the
regulator brane.  Finally the correspondence describes the response of the
dynamics at the UV boundary on the D0`s motion in the bulk via the attached
strings. We have identified above the second time corresponding to
$H_2=TrL_{cal}^2$ as the coordinate scale factor but all even times are
frozen in KdV evolution so the nontrivial evolution starts only due to the
third time. As for the second time there is the additional constraint for the
KdV evolution that it has the "locus configuration" as the initial condition
\cite{poles} which means that it is static configuration of the Calogero
particles with respect to $H_2$. Therefore the KdV - Calogero relation
tells that the evolution has to start at the fixed points of the RG flows.

Let us turn to more interesting case of the  KP-Calogero relation. Now  all
times are relevant and the corresponding elliptic KP Lax operator looks as
follows
\be
(\partial_y +\partial_x^2+2\sum_{i}\wp((x-x_{i}(t,y))\Psi(k,x,y,t)=0
\ee
The flow with respect to $TrL_{2}^2$ corresponds to the  RG flow in SYM
theory so the KP dynamics unlike the KdV one covers the whole parameter space.
The expression for the Calogero Lax fermions can be
read off from the pole expansion of the  BA function of the elliptic
KP solutions
\be
\Psi(k,x,y,t)=\sum \kappa_{i}(t,k,y)\Phi(x-x_{i}(t,y))\exp(kx+k^2y)
\ee
The Lax Calogero equation can be written in terms of $N\times N$ matrixes
\be
L_{Cal}(k)\phi(k)=\lambda \phi(k)
\ee
where $\phi_{i}(k)=\kappa_{i}(k)$.

The correspondence has been previously established only for a fixed value
of the coupling constant, i.e. regulator mass. Therefore we have to
generalize it for the arbitrary  mass. To this end let us think
about the expression for the tau-function as the product of the massless fermion
Green functions on the torus with the modulus
$\tau=\frac{i}{g^2}+\theta$.
The generalization now involves the product of the Green functions of the
massive fermions $S(x-y|m,\tau)$ which can be identified as the regulator ones.
The expression for the $\tau$ function can be interpreted as an energy-stress
tensor generated by the $N_{c}$ sources, whose coordinates $x_i(t)$ are
functions of the RG scale.  The KdV-Calogero relation claims that the RG
dynamics of the D0 branes in the bulk and boundary gravity have to be
correlated.

The arising picture is reminiscent to the Nahm
decription for the monopoles in the brane terms \cite{diac} which
generalizes the analogous interpretation of the ADHM construction
for instantons \cite{doug}.
Nahm matrixes $T_i$ appear as the gauge connection
on D string world volume and Nahm equations themselves
\be
\frac{dT}{ds}=[T,A]
\ee
where s is the coordinate along the D string,
can be interpreted as the
condition for the supersymmetry of the vacuum state in D string
sigma model.

We have seen before that the Lax fermions in the Calogero system,
which can be degenerated to periodic Toda case, correspond to the
localized fermions in the KdV-KP Lax operators. Let us explain
that this is natural in the Nahm picture.
Toda Lax operator can be related with the combination of
the Nahm matrices of a circle of $N_c$ monopoles \cite{sut}.
Hence the time evolution of the Lax fermions in Toda case can be compared
with the following equation for the fermion zero modes known in the Nahm
transform \cite{cg}
\be
(\frac{d}{ds}+\sigma(x+T))\kappa(s,x)=0.
\ee
This formula
also has interpretation in the brane language \cite{diac} if one
considers the probe  D1 string in the background of the brane
configuration which represents monopoles. Then the fermions $\kappa_{i}(s,x)$
can be identified with the fermions on the probe worldvolume. We can
immediately recognize similar interpretation in our case if we treat
regulator branes as the probes.  This matches  the relation between
Calogero-Toda Lax fermions and the localized modes of the regulator fermions.

Having Nahm picture in mind we can suggest that the ends of the
strings attached to D0 branes behave like
a vortices on the regulator brane world volume
and can  be connected to the degrees of freedom in the Calogero-Toda
type dynamics. Equation of motion in the Lax form can be identified with
the Nahm equations and therefore can be considered as the condition of the
unbroken SUSY for the whole configuration. A key point is that the regulators
induce the interaction between the instanton degrees of freedom governed
by the Hamiltonians for Calogero-Toda type systems.

To complete this section let us show that the picture described is consistent
with the "gauge" description of the integrable many-body systems \cite{gn}.
It was demonstrated that the elliptic Calogero dynamical system can be formulated
as the specific 3d topological theory
with the group SL(N,C) for  N particle system defined on the $T^2\times R$
perturbed by the Wilson line in "time" direction in the peculiar representation.
Modulus of the torus coincides with the gauge theory modulus and the phase
space of the model involves the pair $(\bar A, \Phi)$ where $\bar A$ is the
antiholomorphic connection on the torus and $\Phi$ is the one-form.
The Calogero model phase space amounts from the pair above if we impose
the holomorphic Gauss law constraint
\be
\bar \partial \Phi +[\bar A, \Phi]=J \delta (z)
\ee
where we have introduced the source $J_{ij}=\nu (1_{ij}-\delta_{ij})$
corresponding to $CP(N-1)$ coadjoint orbit.

Let us choose the gauge  $\bar A=diag(x_1,...,x_n)$, then
the field $\Phi(z)$ appears to coincide with the Lax operator of
the elliptic Calogero model. In the previous sections we have seen that
coordinates $x_{i}$ can be attributed to the ends of the
strings on the regulator`s world volume hence to get the link
with the discussion above we just have
to perform T-duality in this "holomorphic" direction on the
torus and the diagonal eigenvalues of the connection become
the coordinates .
Let us note that since coordinate $x$ is
parallel coordinate of the regulator branes  they
have to be localized on the initial torus before T duality. This
matches the treatment of the regulator branes in terms
of the localized Wilson line on the bare torus.

One more argument along the same line looks as follows. If one
considers the Hamiltonian formulation of the gauge system with
the Hamiltonian $Tr\Phi^2$ and impose Gauss law with $A_0$ as a
lagrangian multiplier then $A_0$  can be immediately recognized
as A operator from (L,A) representation of the integrable system.
Let us represent a Wilson line above as a path integral
with the action
\be
\int dt(\bar z_i\partial_t z_i +A_0^{ij}\bar z_i z_j)
\ee
where the variables $z_i$ represent coordinates on CP(N-1). Equations
of motion for $z_i$ coincide with the time evolution of the BA
function if we identify $z_i$ as its components. On the other hand
these variables   can be attributed to the regulator degrees
of freedom indeed since the coadjoint orbit in the gauge description
amounts from the massive adjoint hypermultiplet.

\section{Relation with the bulk/boundary issue}

In this section we will try to make additional comments on
the relation
between the  gravity in AdS spaces and integrability
of the dynamical systems on the hidden Higgs branch.
As we mentioned before the natural IIA brane configuration
necessarily includes nonperturbative contributions coming
from the  D0 branes with attached strings.
Therefore the expected associated "nonperturbative"
AdS factors are $AdS_3$ or $AdS_2$.
It appears that both of them are quite
relevant to description of the integrable dynamics.

The idea is to use the fact that both $AdS_3$
and $AdS_2$
geometries admit description purely in terms of
the gauge theory  (see \cite{ads2,ads3} and references therein).
The same is true also for the
dynamical systems of Calogero-Toda type. We would like
to show that these gauge descriptions are in agreement
supporting  the possible interpretation of the
Toda-Calogero dynamics in terms of gravity solutions.

Let us consider first the trigonometric version of the two-body Calogero
system. To this end let us look at the $AdS_3$ gravity
which can be written as SL(2,C) Chern-Simons theory. $AdS_2$
gravity appears as the near the horizon geometry in $AdS_3$
and can be formulated as the SL(2,R) 2d YM theory. It is useful
to consider the Calogero type systems in the Hamiltonian
reduction approach \cite{gn}. Along this approach trigonometric
version can be described equivalently as the geodesic motion on
the cotangent bundle to the group manifold  $T^{*}G$ with the
fixed value of the angular momentum or as the dynamical system
on the cotangent bundle to the affine algebra in the gauge
invariant sector of the Hilbert space.

Since $AdS_3$ coincides with SL(2,R) group manifold we prefer the first
description. The cotangent bundle is the pair
(g,Q) subject to the moment map constraint
\be
g^{-1}Qg-Q=J
\ee
with some level of the moment map at the r.h.s. Therefore the
manifold $T^{*}AdS_3$ plays the role of the unreduced phase
space for the two-body trigonometric Calogero system. After
the reduction proceedure near the horizon
we naturally obtain the structures
related with $AdS_2$.

Geometrically $AdS_2$ is the one-sheet
hyperboloid
\be
x_{1}^2+x_{2}^2-x_{3}^2=R^2
\ee
and its boundary consists of two disconnected lines.
Let us consider the way how the sin-Gordon
solitons could appear in $AdS_{2}$ case. The metric
can be written in the form
\be
ds^2=-sin^2qdt^2 +cos^2qdx^2
\ee
and to yield the space with the constant negative curvature  variable
$q$ has to obey the sin-Gordon equation
\be
-\partial^2_tq+\partial^2_xq =m^2sinq
\ee
with the cosmological constant $m^2$.
To specify
particular solution to the sin-Gordon equation one has to
impose some boundary conditions. The trivial ones
correspond to the pure $AdS_2$ case while the soliton
solution corresponds to the black hole in $AdS_2$ geometry \cite{gk}.
The mapping between soliton parameters and black hole ones
looks as follows;
black hole metric is
\be
ds^2=(m^2r^2-v^4)dT^2 +(m^2r^2-v^4)^{-1}dr^2
\ee
where $v$ is the asymptotic soliton velocity,
the mass of the black hole is $M=\frac{v^2}{m}$ and the
ADM energy coincides with the
nonrelativistic kinetic energy of the
soliton.  Note that the rational case has been
discussed recently in the similar context in \cite{gt}.
This identification of the sin-Gordon solition with the
$AdS_2$ black hole suggests the relation of the many-body
trigonometric Calogero system with the multi-black hole
solution in $AdS_2$ since it governs the n-soliton
sector in sin-Gordon model.

We have observed at the
boundary the KdV Lax operator
and the singular energy-stress tensor T(x) with
double poles at $x_{i}$ which were identified with the ends
of strings attached to the bulk D0 branes.
It is wellknown that KdV equation can be considered as the perturbation
of the Liouville action so we have to check that our
picture is consistent with this fact.
Since Liouville
theory lives at the boundary of the $AdS_2$ and $AdS_3$
spaces one might be interested in the geometrical
meaning of such T(x). In turn it is useful to remind
that the Liouville theory can be defined as the geometrical
action on the coadjoint Virasoro orbit with zero
representative \cite{as}. However there are several types of the
Virasoro coadjoint orbits and there are
so called special coadjoint orbits which are of interest
here. The very existance of these orbits
amounts from the nontrivial $\pi_{1}(SL(2,R))$ providing
the topologically nontrivial gauge configurations
which after the hamiltonian reduction yield special
coadjoint orbits. The key feature concerning this type of
orbits is that there is no constant two-differential
representative on the orbit and they
necessarily have some number of poles
reflecting the fact that the corresponding vector field
stabilizer has the same number of zeros.
The generalization of the Liouville action for the special
orbits has been found in \cite{special} and it was shown
that it corresponds to the Liouville field interacting
with the finite number of the quantum mechanical degrees of
freedom which can be treated as the special solutions
coming from the vortex gauge configurations. Therefore we
indeed have the Liouville action perturbed by the finite
number of degrees of freedom which is consistent with  the
appearance of KdV solutions with the finite number of moduli.

Let us discuss now  the possible interpretation of the
formulae for the second derivative of the prepotential in the N=2 SYM theory
with respect to the Whitham times. These formulae can
be derived both within the Donaldson \cite{don} and Whitham
approaches and have the following structure
\be
\frac{\partial^2{\cal F}}{\partial T^m\partial T^n}
= -\frac{\beta}{2\pi i} \left({\cal H}_{m+1,n+1}
+ \frac{\beta}{mn}\frac{\partial {\cal H}_{m+1}}{\partial a^i}
\frac{\partial {\cal H}_{n+1}}{\partial a^j}
\partial^2_{ij} \log \theta_E(\vec 0|{\cal T})\right)
\label{2der}
\ee

We will focus below on the derivatives  over
first time $T_1=\log \Lambda$ which provides the expression
for the correlator $<Tr\Phi^2(x) Tr\Phi^2(y)>$ in terms
of the $\theta$ functions on the spectral curve.

Let us show that  this expression in reminiscent with the
recipe \cite{gkp,w1} for calculation of the quantum YM correlators
in terms of the classical solutions in supergravity with
the appropriate boundary value of the corresponding source field.
Calculation of the "action on the supergravity solution" is a subtle
point in the nonconformal case. It goes as follows. At the first step
one has to consider the dynamics of the "fast" degrees of freedom which
we identified with the small size instantons. The corresponding
solution of the  dynamical equations for the fast degrees of freedom is
encoded by the spectral curve which is invariant of the AdS "radial time"
evolution. To recognize prepotential $\cal F$ as the action on the gravity
side we have to average over the fast dynamics admitting the slow evolution on
the Coulomb branch.  This slow dynamics is governed by the Whitham system and
the prepotential is effective action for this evolution indeed \cite{gkmmm}.

In the brane terms these two evolutions
looks as follows. Fast Calogero-Toda dynamical system
involves the motion of D0 branes with positions of D4 branes
being fixed. Whitham dynamics arises from the dynamics of D4 branes
themselves. The evolution with respect to the  Whitham
time $T_1$ represents a
motion of D4 branes which
for SU(2) can be  identified with the Gurevich-Pitaevskii
solution \cite{gkmmm}. This type of evolution can be  treated
as the condition of the approximate invariance of the spectral curve under
the RG flow. Indeed the very meaning of the Whitham dynamics
is to provide the validity of the classical equation of
motion in the unperturbed theory if the first order
perturbative correction  is taken into account. The perturbation
now is identified with the motion of D4 branes and this
condition imposes the strong restriction on their evolution.

According to the general prescription to obtain some correlator
one has to calculate
the action on the classical solution in gravity
and differentiate it over the boundary value
of the corresponding "source". In the N=2 case there are
two natural boundaries: UV and IR ones and we would like
to get the correlators at UV scale at first. Then using
the fact that the spectral curve is the "RG invariant"
we can use it to derive the correlators in the IR.
On the general grounds one might expect that it would be
correct only if the anomalous correlators are saturated
by the zero modes.
Since we consider
correlators of operators ${\cal O}_2=Tr \Phi^2$, the boundary
value of the dilaton field is relevant.
Hence
the variation of the action on the solution with respect
to $T_1=\log \Lambda=\log M +\frac {i\beta}{g^2(M)}$
coincides with the variation over the UV boundary
dilaton value in agreement with the general recipe.

It is also instructive to consider  correlator in terms
of the fermionic Green functions on the spectral curve \cite{gmmm}
\be
<Tr\Phi^2 Tr\Phi^2> \propto \frac{\partial u}{\partial a} \int
\Psi_{E}^{2}(\lambda)
\ee
where $\Psi_{E} (\lambda,\infty)$ is the Szego
kernel with the even theta-characteristic E on the Toda chain spectral curve
and the integral is evaluated along A-cycle.  The variation over $T_1$ can be
considered as the variation with respect to the regulator mass hence we
immediately derive the fermionic bilinears on the worldvolume of the
regulator brane.  Therefore we have the product of 2D fermionic Green
functions from the very beginning. However the Green function on the spectral
curve enters the answer so it appearence has to be explained.
The possible explanation concerns the transition to the
momentum space which is intimately related to the spectral curve.
Indeed the spectral curve appears as the Fermi surface for the Toda Lax
fermions which on the other hand were identified with the localized
topological modes of the regulator fermion at the end of the  string. So the
spectral curve is the Fermi surface for this regulator modes too and the
structure of the answer provides some evidence for the conjecture that the
correlator is saturated purely by the topological "zero" modes of the
fermionic component of the regulator superfield.

\section{Discussion}

In this paper we suggest the RG intepretation of the integrable
dynamics governing the low-energy behaviour of  N=2 SUSY YM
theories. It was conjectured that both dynamical systems of
Hitchin-spin chains and Whitham types can be considered as the
RG equations.
It seems that the UV cutoff introduced to regularize theory
induces nontrivial consequences for the small instanton dynamics.
Namely it generates the interaction between the small size instantons
which also has some back reaction on the regulator degrees of freedom.

The time in the Hitchin dynamical system in this scenario can be identified
with the space RG scale and the integrals of motion in
the Hitchin type systems
are the invariants of the RG evolution. We tried to argue that the Lax
fermions can be attributed to the regulator degrees of freedom. The very
meaning of the integrability is to provide the regularization of the
nonperturbative effects consistent with the RG flows.  The dynamics of D0
branes in the bulk should be correlated with the dynamics of the degrees of
freedom at the regulator scale. Let us emphasize that we avoided the
specific description of the "regulator branes" since the pure SYM low-energy
effective action can be derived from the different theories. For instance
regulators can be in  different representation of the matter group
so the  pure gauge low-energy theory forgets about the brane realization of the
regulators when their masses tend to infinity.

The RG interpretation shed some light on the
possible role of the quantization of the integrable
systems in the SYM setup. Indeed  the stringy point of view
suggests that to quantize the RG flow one has at first
to consider the gravitational dressing and then the special
pinched world sheet geometries have to be taken into account \cite{sch}.
One more possible speculation is that the space scale is
gravitationally dressed in the momentum space like the
space cutoff $M^{-1}$ is translated into the
"gravitationally dressed" cutoff
\be
\Lambda= M\exp\frac{iN_{c}}{g^2}=M\exp(<N_{c}\exp \phi>).
\ee
This simple argument indicates that transition from the
coordinate to the momentum representation necessarily involves
the account of the gravity.

The brane geometries for N=1 SYM and for
softly broken N=1 theory  are known so some analogous
treatment possibly can be developed. In fact  zero modes of the regulators
were used in \cite{nsvz} to derive the perturbative $\beta$ function
in the N=1 SYM theory. It seems that to evaluate the full $\beta$ function
the complicated  structure at the regulator scale has to be taken into account.
Let us also remark that some  integrability behind the RG flows in d=4 N=1
theory can be
expected since the RG behaviour of its N=2 d=2 counterpart is
governed by the integrable 2d Toda  lattice \cite {vafacec}.

I would like to thank H.Leutwyler for the hospitality at the Institute
for Theoretical Physics at Bern University, while
part of this work was done.
I am grateful to A.Mironov and A.Morozov for the useful discussions and
C.Schmidhuber for the comments on the gravitational
dressing of the RG flows. The work
is supported in part by grants INTAS-96-0482, RFBR 97-02-16131
and Schweizerischer Nationalfonds.

\end{document}